\documentclass[12pt]{article}

\usepackage{graphicx}
\usepackage{color}

\begin{document}
\baselineskip=25pt

\vspace{0.5cm}

\begin{center}
{\bf Physical realization and possible identification of topological excitations in quantum Heisenberg anti-ferromagnet on a two dimensional lattice}
\end{center}
\begin{center}
\vspace{0.5cm}
{ Ranjan Chaudhury}$^a$ and { Samir K. Paul}$^b$\\
$S.~ N.~ Bose~ National~ Centre~ For~ Basic~ Sciences , ~~ Block$-$JD,~ Sector$-$III , ~ Salt~ Lake$\\
$Calcutta$-$700098,~India$
\end{center}
\vspace{0.5cm}

a)    {$ranjan_{021258}@yahoo.com$ , ranjan@boson.bose.res.in}

b)    { smr@boson.bose.res.in}

\vspace{1.00cm}

Physical spin configurations corresponding to topological excitations , expected to be  present in the XY limit of a quantum spin $\frac {1}{2}$ Heisenberg anti-ferromagnet, are probed on a two dimensional square lattice .  Quantum vortices (anti-vortices) are constructed in terms of coherent staggered spin field components, as limiting case of meronic  ( anti-meronic ) configurations . The crucial role of the associated Wess-Zumino-like ( WZ-like ) term is highlighted in our procedure .  The time evolution equation of coherent spin fields used in this analysis is obtained by applying variational principle on  the  quantum Euclidean action  corresponding to the  Heisenberg anti-ferromagnet  on lattice . It is shown that the WZ-like term can distinguish between vortices and anti-vortices only in a charge sector with odd topological charges. Our formalism is distinctly different from the conventional approach for the  construction of quantum vortices ( anti-vortices ) .

\vspace{0.2cm}

{\bf Key~Words} : Wess Zumino, Topological, Heisenberg, Vortices.
\vspace{10.00cm}

{\bf 1. Introduction}\\

\vspace{0.1cm}
  One of the most convenient approaches to investigate  the origin of topological excitations in quantum  spin systems is through the  coherent state formulation [1-5]. The geometrical phase ( Berry phase ) which arises in the partition function of the quantum spin models in this approach ,  plays the central role in deciding the topological sectors of various spin field configurations .  For instance, in the case of the Heisenberg antiferromagnetic  chain in the long wave length limit, one has a two-dimensional ( one spatial plus one pseudo-temporal) nonlinear sigma model . This model contains  the  geometrical phase which acts as the topological WZ term [1] . For lattices with  spatial  dimensions of two or more, such type of term vanishes in  the long wavelength limit [4]. Interestingly  enough, this term is found to survive in the medium wave length limit and finds its signatures in various numerical and analytical calculations, as well as in neutron scattering experiments [6 - 9].  It may be recalled that in the long wavelength limit, one retains terms up to first order in spatial derivatives in the expansion of the spin field within the WZ-term. In this limit, the WZ-term survives in the case of a spin chain   but vanishes for a square lattice [2,4].  Therefore in the latter case, it is necessary to retain terms up to 2nd order in spatial derivatives to extract a non-vanishing contribution. This requires a total length scale of 4 lattice spacings (viz., $2a$ in the positive direction and $2a$ in the negative direction, $a$ being the lattice parameter ) along both the axes. This translates in the $\bf q$ space as a regime around $(|{\frac {\pi}{2a}}| , |{\frac {\pi}{2a}}|)$ .  Thus we are in the vicinity of the mid-zone region in the q-space . Calculations based on this region are referred to  as "medium wavelength approximation". Hence it is  extremely important  to perform calculations explicitly on the lattice in order to get a finite contribution to the WZ-term in the lowest order .  The geometrical phase in its discrete form is the lattice analogue of the Wess-Zumino (equivalently Berry phase) term. Henceforth we refer to this discrete expression as "WZ-like" term .\\    
One of the major motivations for  our present study  has been theoretical determination of the dynamical structure factor $S({\bf q},\omega )$ for low dimensional quantum antiferromagnets, particularly in view of the inelastic neutron scattering experiments  on the cuprates [6] . In particular,  the occurrence of 'central peak' in the experimentally observed $S({\bf q}, \omega )$  in the medium wavelength regime, strongly indicates possible existence and dynamics of topological excitations [6,7].  The 'central peak' refers to the peak occuring at $\omega = 0$, in the plot of $S({\bf q}, \omega )$ versus  $\omega $  in the constant $q$-scan. Generally, this is an important signature for the spin dynamics driven also by the translational motion of the topological excitations and defects [3, 7].  Furthermore , the spin-spin correlation length $\xi (T)$ as calculated from the experimentally extracted $ {C({\bf q})} $ ( the static q-space spin-spin correlation function ) for spin $ {\frac {1}{2}}$ two-dimensional quantum Heisenberg anti-ferromagnet (QHAF), shows marked  departure  from the renormalized spin wave theory, in the cuprates [9,10] . \\  
These observations combined with the theoretical analysis described above, suggest that by retaining the discrete lattice structure for low dimensional quantum antiferromagnets it should be possible to get a nonvanishing contribution of the Berry Phase or the WZ-like term [3,8,9 ] . \\  
Earlier detailed analysis was done by us for a general quantum anisotropic Heisenberg spin model using coherent state formalism in a quasi-continuum limit [3,8]. Now  we deal with the extreme quantum case of a strongly anisotropic  i.e., XY-limit of a spin $\frac {1}{2}$ QHAF explicitly on a lattice.\\
~~~ It is worthwhile to refer to another approach to study topological excitations [9] . In that approach the technique of Schwinger Boson Mean Field Theory (SBMFT) was applied on a quantum Heisenberg anti-ferromagnetic spin system at an intermediate length scale , assuming that the corresponding field theoretic action does not contain any Berry phase$/$ Wess Zumino ( topological like )term , in contrast to the real situation. Therefore the topological excitations are introduced  heuristically there [9] . \\   
 ~~~~~~The main aim of this paper is to verify the presence of the WZ-like term and examine the properties of the physical spin configurations . We demonstrate that the WZ-like term really identifies a large class of the  topological excitations and clearly  differentiates between quantum vortices  and antivortices  with different charges . \\

{\bf 2.~Mathematical Formulation}\\
{\bf 2.1 ~Action and the WZ-like term }\\
  The quantum Euclidean action ${{\mathcal S}_E}$ for the  coherent spin fields
${\bf n}({\bf r},t)$ can be written as [2] \\
\begin{equation}
{{\mathcal S}_E} = -is g{\sum_ {\bf r}}   {{\mathcal S}_{WZ}}[{\bf m}({\bf r},t)] + {\int_0^{\beta}}dt {\mathcal H}({\bf n})
\end{equation}
where '$s$' is the magnitude of the spin ($s = {\frac{1}{2}}$ in the present case ) , $\bf r$ is the position vector of a lattice site and 
\begin{eqnarray}
\langle {\bf n}\vert {\bf S}\vert {\bf n}\rangle & = & s{\bf n} \nonumber\\
{\mathcal H}({\bf n}) & = &\langle {\bf n}\vert {\mathcal H}({\bf S})\vert {\bf n}\rangle  \nonumber\\
\vert {\bf n}\rangle & = &  {\prod_{\bf r} } \vert {\bf n}({\bf r},t) \rangle  \nonumber\\
\end{eqnarray}
where $\prod$ denotes the direct product of all coherent spin states over the spatial lattice . The quantity  ${\mathcal H}({\bf S})$ is the spin Hamiltonian on the lattice and ${{\mathcal S}_{WZ}}[{\bf m}({\bf r},t)]$ is the WZ-like term on a single lattice site.\\     
${\mathcal H}({\bf S})$ in our calculation is given by equation (5)   in  spin $s$  representation and  ${\mathcal H}({\bf n})$ the corresponding Hamiltonian in the coherent spin fields  ${\bf n}({\bf r},t)$ ( see equation (6) ).  The variable $t$ denotes the pseudo-time  (Euclidean time) appropriate to  the above fields and has  the dimension of inverse temperature .  The parameter  $\beta$ stands for $ {\frac{1}{kT}}$ as usual;  $T$ being the real thermodynamic temperature of the spin system. The WZ-like term $S_{WZ}$ corresponding to a single spatial lattice point ${\bf r}$ at a fixed time 't' is given as follows [2]:
\begin{equation}
{{\mathcal S}_{WZ}}[{\bf m}({\bf r},t)]={\int_0^{\beta}}dt{\int_0^1}d{\tau} {\bf m}({\bf r},t,{\tau})\cdot {\partial_t}{\bf m}({\bf r},t,{\tau})\wedge {\partial_{\tau}}{\bf m}({\bf r},t,{\tau})
\end{equation}
with ${\bf m}({\bf r},t,0)\equiv {\bf n}({\bf r},t)$, ${\bf m}({\bf r},t,1)\equiv {{\bf n}_0}({\bf r})$, and
${\bf m}({\bf r},0,{\tau})\equiv {\bf m}({\bf r},\beta ,\tau )$, $t\in [0,\beta ]$, $\tau \in [0,1]$ .\\
The expression in equation(3) is the area of the cap bounded by the trajectory  $\Gamma$ parametrized by ${\bf n}({\bf r},t)$ $[{\equiv} ({n_1}({\bf r},t),{n_2}({\bf r},t),{n_3}({\bf r},t))]$  on the sphere:
\begin{equation}
{\bf n}({\bf r},t)\cdot {\bf n}({\bf r},t) = 1
\end{equation}
Furthermore the fields ${\bf m}({\bf r},t,{\tau})$ are the fields in the higher dimensional $(t, \tau)$-space and the boundary values ${\bf n}({\bf r},t)$ are the coherent spin fields . The field ${{\bf n}_0}({\bf r})$  is the fixed point $(0,0,1)$ on the above sphere and the  state vector $\vert {\bf n}({\bf r},t) \rangle$ appearing on the right hand side of equation (2), is the spin coherent state at a single lattice point $\bf r$ [1-3] .\\
 For the  two dimensional lattice , we express ${\bf n}({\bf r},t)\equiv {\bf n}(ia ,ja , t)$ as   ${\bf n}(ia ,ja )$ for breavity  .\\

{\bf 2.2 ~The Spin Hamiltonian }\\
The spin Hamiltonian corresponding to anisotropic Heisenberg spin system of $XXZ$ type with antiferromagnetic coupling is given by
\begin{equation}
{\mathcal H}({\bf S}) =  g{\sum_{\langle {\bf r},{\bf r\prime}\rangle}}{\bf {\tilde S}}({\bf r})\cdot {\bf {\tilde S}}({\bf {r\prime}}) + g{\lambda_z}{\sum_{\langle {\bf r},{\bf {r\prime}}\rangle}}{S_z}({\bf r}){S_z}({\bf {r\prime}})
\end{equation}
with $g > 0$ and $0\le {\lambda_z} < 1$ , $\bf r$,$\bf {r\prime}$ running over the lattice, and $\langle {\bf r},{\bf {r\prime}}\rangle$ signifies nearest neighbours  and ${\bf S} = ({\bf {\tilde S}}, S_z)$.  Here ${\bf {\tilde S}}\equiv {({S_x},{S_y})}$  is the projection of the operator $\bf S$ onto the XY-plane . \\
It follows from equations $(2)$ and $(5)$ that the spin Hamiltonian in terms of coherent spin fields is given by
\begin{equation}
{\mathcal H}({\bf n}) = g~{s^2}{\sum_{\langle (i,j),({i^{\prime}} ,{j^{\prime}})\rangle}}{\bf {\tilde n}}(ia,ja)\cdot {\bf {\tilde n}}({i^{\prime}}a ,{j^{\prime}}a) + g~{s^2}{\lambda_z}{\sum_{\langle (i,j),({i^{\prime}} ,{j^{\prime}})\rangle}}{n_3}(ia,ja) {n_3}(ra,sa)
\end{equation}
Here ~~ ${\bf {\tilde n}}(ia,ja)\equiv {({n_1}(ia,ja),{n_2}(ia,ja))}$ is the projection of the coherent spin field ~~$\bf n$~~ onto the XY-plane . Mathematically , ${\bf {\tilde n}}(ia,ja) = \langle {\bf n}\vert {\bf {\tilde S}}(ia,ja)\vert {\bf n}\rangle$ ~~where the state vector $\vert {\bf n}\rangle$ is given by equation (2)    .\\ 
From equations (1), (2)  and  (6) the quantum action (Euclidean)for the two dimensional anisotropic spin system with antiferromagnetic coupling is given by  [2]
\begin{eqnarray}
{{\mathcal S}_E}& = &-is{\sum_{i,j}}{{\mathcal S}_{WZ}}[{\bf m}(ia,ja)] + {\int_0^{\beta}}dt [~g~{s^2}~{\sum_{\langle (i,j),({i^{\prime}} ,{j^{\prime}})\rangle}}{\bf {\tilde n}}(ia,ja)\cdot{\bf {\tilde n}}({i^{\prime}}a ,{j^{\prime}}a) \nonumber\\
&  & + g{\lambda_z}{s^2}{\sum_{\langle (i,j),({i^{\prime}} ,{j^{\prime}})\rangle}}{n_3}(ia,ja){n_3}({i^{\prime}}a ,{j^{\prime}}a)]
\end{eqnarray}
where the constraint given by equation (4) has to be satisfied at each lattice point .\\
As our system is a quantum antiferromagnet , we assume the spin configurations  to have correlations exhibiting  a bipartite symmetry , when the temperature is not too high [8]. We therefore stagger the configuration as:
\begin{equation}
{\bf n}(ia,ja) \longrightarrow {{(-1)}^{i+j}}{\bf n}(ia,ja)
\end{equation}
Thus we obtain from equations (4) , (7) and (8) the following total quantum Euclidean action for the anisotropic Heisenberg antiferromagnet
\begin{eqnarray}
{{\mathcal S}_E^{stagg}}& = &-is{\sum_{i,j}}~{{(-1)}^{i+j}}~{{\mathcal S}_{WZ}}[{\bf m}(i.a,ja)] + {\int_0^{\beta}}dt~ g~{s^2}{\sum_{\langle (i,j),({i^{\prime}} ,{j^{\prime}})\rangle}} [~{{(-1)}^{i+j+{i^{\prime}}+{j^{\prime}}}}                      {\bf {\tilde n}}(ia,ja)\cdot{\bf {\tilde n}}({i^{\prime}}a ,{j^{\prime}}a) \nonumber\\
&  & +~~g~{\lambda_z}{s^2}{\sum_{\langle (i,j),({i^{\prime}} ,{j^{\prime}})\rangle}}{{(-1)}^{i+j+{i^{\prime}}+{j^{\prime}}}}~           {n_3}(ia,ja){n_3}({i^{\prime}}a ,{j^{\prime}}a)] \nonumber\\    
&  & - {\int_0^{\beta}}~dt~{\sum_{i,j}} {a^2}{\lambda_{i,j}} [{{\bf n}^2}(ia,ja) - 1] 
\end{eqnarray}
Hence for the nearest neighbour interaction the above action is given by:
\begin{eqnarray}
{{\mathcal S}_E^{stagg}}& = &-is{\sum_{i,j}}~{{(-1)}^{i+j}}~{{\mathcal S}_{WZ}}[{\bf m}(i,j)] + {\int_0^{\beta}}dt \{~-g~{s^2}{\sum_{i,j}}[{\bf {\tilde n}}(i,j)\cdot{\bf {\tilde n}}(i+1,j) \nonumber\\
&  & + {\bf {\tilde n}}(i,j)\cdot{\bf {\tilde n}}(i,j+1)~] ~-~g~{\lambda_z}{s^2}{\sum_{(i,j)}}[{n_3}(i,j){n_3}(i+1,j) + {n_3}(i,j){n_3}(i,j+1)] \nonumber\\
&  & - {\int_0^{\beta}}~dt~{\sum_{(i,j)}}~{a^2}{\lambda_{i,j}} [~{{\bf n}^2}(i,j) - 1] \}
\end{eqnarray}
The last term on the right hand side of the above equation is the lattice version of the term ${\int} {d^2}x {\int_0^{\beta}} {\lambda} ({\bf x}, t)({{\bf n}^2}({\bf x},t) - 1)$ where ${\lambda}({\bf x},t)$  is an auxiliary field playing the role of a multiplier . The equations of motion which follow from the  minimisation of the above action are given by equations (A.1) in the Appendix . The purpose of this exercise  is to obtain an  expression for  ${{\mathcal S}_{WZ}^{stagg}}$ given by equation (11) below , in terms of individual contributions from lattice sites. The quantity ${{\mathcal S}_{WZ}^{stagg}}$ is topological-like in the sense that the $WZ$ term survives in the medium wave length limit and it indicates topological excitations [1, 8-10] . One can derive the expression for ${{\mathcal S}_{WZ}^{stagg}}$ by using equations  (11)-(13) below ( for example , as we have  done  in  the case of 1-vortex ;resulting in equation (A.5) in the Appendix ) . Then we substitute for $\partial_t {\bf n}$ from the equations of motion ( $A.1$ ) into the expression for ${{\mathcal S}_{WZ}^{stagg}}$ ( in equation ( A.4 ) ) .    \\
The first term on the right hand side of the above Eqn. (10) is the $WZ$-like  term on the lattice which  is formally written as:\\
\begin{equation}
{{\mathcal S}_{WZ}^{stagg}} = {\sum_{i,j}}~{{(-1)}^{i+j}}~{{\mathcal S}_{WZ}}[{\bf m}(i,j)]
\end{equation}
 It is  possible to evaluate only the 'difference' of ${{\mathcal S}_{WZ}}[{\bf m}(ia,ja)]$ terms from two neighbouring lattice sites, in terms of the coherent spin fields ${\bf n}(ia,ja)$. Thus to extract the topological-like contribution from  ${{\mathcal S}_{WZ}^{stagg}}$, we use the following expressions for the above 'difference'  [2]  :
\begin{eqnarray}
{\delta_x}{{\mathcal S}_{WZ}}[{\bf m}(\bf r)] & \equiv & {{\mathcal S}_{WZ}}[{\bf m}(ia,ja)] - {{\mathcal S}_{WZ}}[{\bf m}((i-1)a,ja)]\nonumber\\
& = &{\int_0^{\beta}}dt [{\delta_x} {\bf n}\cdot ({\bf n}\wedge {\partial_t}{\bf n})]({\bf r})\nonumber\\
{\delta_y}{{\mathcal S}_{WZ}}[{\bf m}(\bf r)] & \equiv  & {{\mathcal S}_{WZ}}[{\bf m}(ia,ja)] - {{\mathcal S}_{WZ}}[{\bf m}(ia,(j-1)a)]\nonumber\\
 & = &{\int_0^{\beta}}dt[{\delta_y} {\bf n}\cdot ({\bf n}\wedge {\partial_t}{\bf n})]({\bf r})
\end{eqnarray}
where
\begin{eqnarray}
{\delta_x}{\bf n}(ia,ja) & \equiv &{\bf n}( ia,ja ) - {\bf n}( (i-1)a,ja )\nonumber\\
{\delta_y}{\bf n}(ia,ja) & \equiv &{\bf n}( ia,ja ) - {\bf n}( ia,(j-1)a )
\end{eqnarray}
 with ${\bf r}\equiv (ia,ja)$ , and $i, j = 1,2,........., 2N$ .\\
From equations (11) - (13) we can write ${{\mathcal S}_{WZ}^{stagg}}$ as given by equation (A.4) in the Appendix .

\vspace{0.5cm}

{\bf 3. Calculations and Results}\\
 We now construct vortices ( anti-vortices ) in the flattened meron configuration limit with staggered spin fields .   It is important to point out at this stage that it is necessary to retain both  $\lambda_z$ and $n_3$ as infinitesimal but nonvanishing quantities , in order to generate a topological character of the above configurations ( see equation (7) and Ref.[3] ). After the construction of an elementary vortex ( anti-vortex ) plaquette with coherent spin field    components  ${n_1}(ia , ja)$, ${n_2}(ia , ja)$ and  ${n_3}(ia , ja)$  we  present calculations for ${{\mathcal S}_{WZ}^{stagg}}$ corresponding to vortices (anti-vortices). We analyse all the higher charged vortices ( anti-vortices ) by decomposing them in terms of  elementary plaquettes of unit charge .\\ 
Our findings are :\\
I)  Odd-charged  configurations can be described consistently in our scheme . \\
II) Even-charged configurations however , behave in an anomolous way , which is explained in the Appendix .\\
III)  Simple calculations based on the Hamiltonian with staggered spin fields[ see equations (10) and (A.6)] show that the excitation energy of a 1-vortex, as measured from the ground state, is approximately $+12g{s^2}$ . It may be noted that the ground state spin configuration is obtained by taking the coherent spin vectors (${\bf n}(ia,ja)$)   aligned in the same direction . These topological excitations can indeed be produced even at very low temperature by means of quantum fluctuations , thereby maintaining a steady vortex-anti-vortex pair density even at temperature close to zero [14] .\\

{\bf 3.1a ~Analysis of 1-vortex} \\
\begin{figure}[!htbp]
\begin{center}
\includegraphics[keepaspectratio,width=7cm]{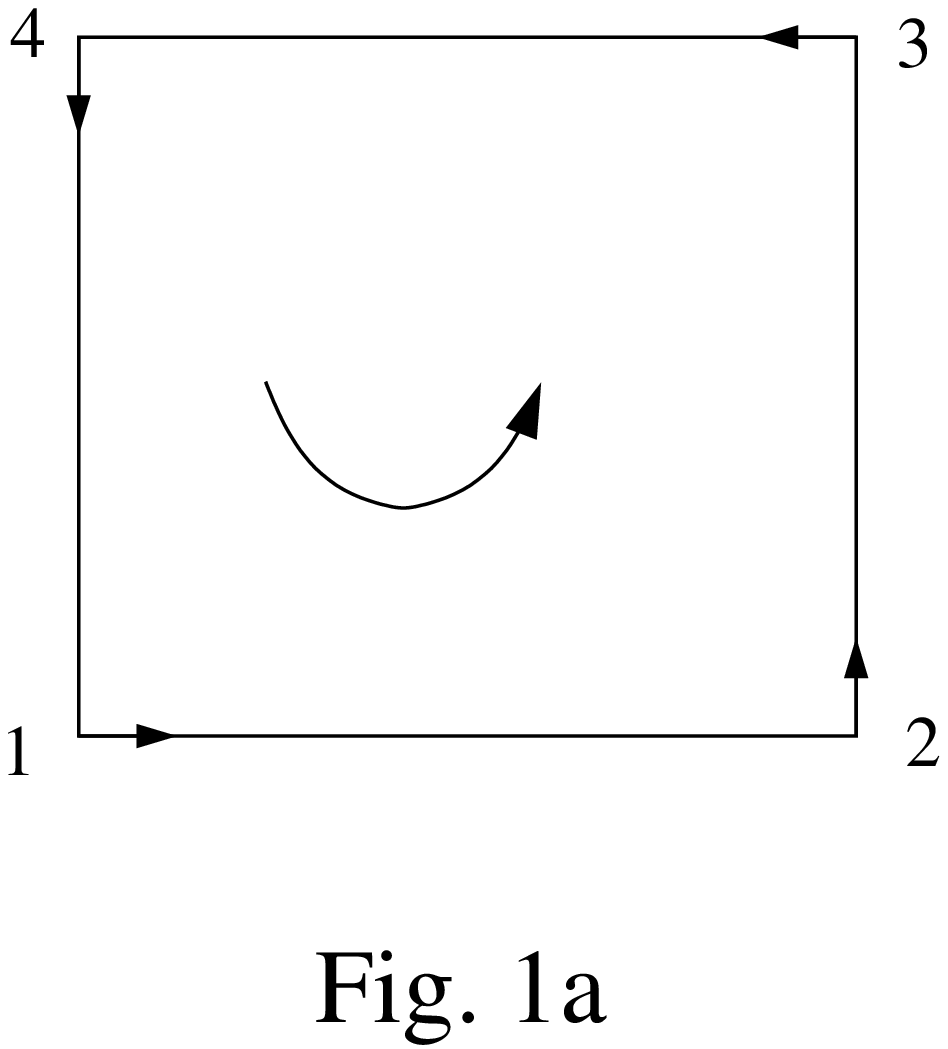}
\includegraphics[keepaspectratio,width=7cm]{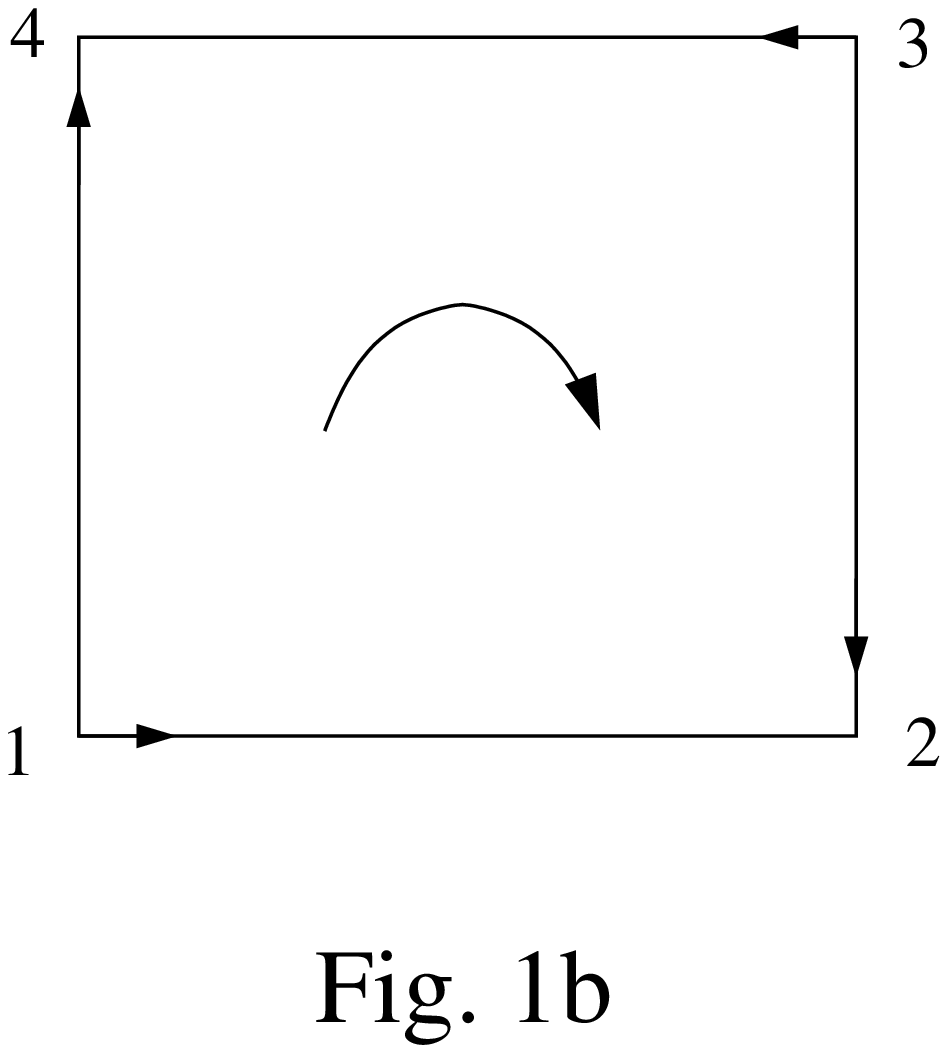}
\end{center}
\caption{(a) 1-vortex , (b) 1-anti-vortex}
\end{figure}

We assign  the coordinates  $( (i-1)a , (j-1)a )$,    $( ia,(j-1)a )$ , $( ia,ja )$  and $( (i-1)a, ja )$  to the  vertices 1, 2, 3 and 4 respectively,  of an elementary 1-vortex [$Fig.1$] plaquette and construct the quantum vortex with staggered spin fields.\\
   As we are interested in the extreme $XY$- anisotropic limit, we  assume ${n_3} ({i^{\prime}}a ,{j^{\prime}}a) =  \sin \epsilon ({i^{\prime}}a ,{j^{\prime}}a)$ at each lattice point $({i^{\prime}}a ,{j^{\prime}}a)$  , where $\epsilon ({i^{\prime}}a ,{j^{\prime}}a) $ is a very small positive quantity dependent on lattice site $({i^{\prime}}a ,{j^{\prime}}a)$. Then it follows from  equation (4)  that  within a vortex (or anti-vortex) the following solutions emerge :\\
  At any vertex , ${n_1}$ = $\pm (1-\delta)$  ; ${n_2}$ = $\pm {\sqrt {[{{\cos}^2}{\epsilon (la,ma)}  - {{(1-\delta )}^2}]}}$ and vice versa .  \\
For illustration [ Fig.-1 ] we have a quantum vortex of charge $+1$ in which the horizontal arrow $\rightarrow$ at a vertex represents  $n_1$ with value $1-\delta$ and the vertical arrow $\uparrow$ implies $n_2$ with value $1-\delta$. Further, in this figure the horizontal arrow $\leftarrow$ at a vertex denotes $n_1$ having value  $-({1-\delta})$ and the vertical arrow $\downarrow$ represents $n_2$ with value $-({1-\delta})$.  For vortex with stable spin configuration  the horizontal or vertical arrow should have a steady magnitude  of  $\pm ({1-\delta})$  . This implies that the  quantity  $\delta$ must be independent of time ( described  in the Appendix ). It may be remarked that the vortex configuration (shown in Fig.1a) has a topological charge $+1$ , as the spin rotates through an angle $+2\pi$ in traversing the boundary once in the anti-clockwise sense.\\
 In this connection let us point out that usually in a two dimensional vortex  corresponding to spin $\frac {1}{2}$ quantum spin model,  the states $\vert \rightarrow \rangle$ , $\vert  \leftarrow \rangle$  are taken to be the eigenstates of $S_x$ with eigenvalues $+{\frac{1}{2}}$ , $-{\frac{1}{2}}$ respectively.  Similarly the  states $\vert \uparrow \rangle$ , $\vert  \downarrow \rangle$ are the eigenstates of $S_y$ with eigenvalues $+{\frac{1}{2}}$ ,   $-{\frac{1}{2}}$ respectively [11] .  However here we are not following that scheme . We make use of the coherent states given by $\vert {\bf n} \rangle = {\cos{\frac{\theta}{2}}}{\vert  {\frac{1}{2}}\rangle} + ({e^{-\phi}}) {\sin{\frac{\theta}{2}}}{\vert  {-\frac{1}{2}}\rangle}$ and evaluate the expectation values of $S_x$ and $S_y$ (referred earlier as $n_1$ and $n_2$ respectively ).In our picture the horizontal and vertical arrows represent these expectation values  $n_1$ and  $n_2$ respectively . To be more precise our configurations are in fact  'flattened merons(anti-merons)', mimicking a  vortex (or anti-vortex) in the limiting  case $\lambda_z \rightarrow 0$ [12, 13]    . \\

  $This$ $prescription$ $will$  $be$  $followed$ $for$ $constructing$ $the$  $vortices$ $( anti-vortices )$ $of$ $higher$ $charge$ $values$ $as$ $well$ .\\

{\bf 3.1b ~Contribution of ${{\mathcal S}_{WZ}^{stagg}}$ }\\
 
Let us now look at the symmetry properties of the contribution of ${{\mathcal S}_{WZ}^{stagg}}$ to the vortex plaquette, which we have denoted by $ {{[{{\mathcal S}_{WZ}}]}_{(1-vortex)}}$ . It is shown that (see Appendix) the contriburion of $ {{\mathcal S}_{WZ}^{stagg}}$  to a vortex plaquette can be decomposed into two parts such that the  first part  remains invariant when  going over from a 1-vortex configuration to the corresponding anti-vortex configuration ; whereas the second part changes sign  under this operation. We denote the first part by $A$ and the second part by $B$. This transformation from vortex to anti-vortex is implemented by changing ${n_2}(2)$ and ${n_2}(4)$ in Fig.1 to -${n_2}(2)$ and -${n_2}(4)$ respectively, in the case of a 1-vortex.Algebraically this means that $A$ contains terms which are  {\bf quadratic or of even degree} in ${n_2}(2)$ and ${n_2}(4)$ while $B$ contains terms that are linear or of odd degree in ${n_2}(2)$ and ${n_2}(4)$. Using equations (11)-(13) we cast ${{\mathcal S}_{WZ}^{stagg}}$  in the form $ A + B$ as explained in the Appendix . \\

 {\bf {We~~adopt~~the~~following~~algorithm~~for~~the~~construction}}\\
{\bf {~~of~~higher~~vortices~~(anti-vortices)}}:-    \\
For the construction of higher vortices we calculate the contribution of  ${{\mathcal S}_{WZ}^{stagg}}$ given by equations (11)-(13) on a plaquette by algebraically adding the contributions of ${{\mathcal S}_{WZ}^{stagg}}$ on  each of the individual elementary plaquettes (subvortices) with a weightage factor of $\frac{1}{2}$ to the common bonds shared between the pairs of adjacent subvortices. We are interested in those field configurations for which the  contributions  of ${{\mathcal S}_{WZ}^{stagg}}$  on the common bonds cancel each other (see Appendix) and only the contribution on the peripheral boundary remains.\\

{\bf 3.2 ~Analysis of 2-vortex }
\begin{figure}[!htbp]
\begin{center}
\includegraphics[keepaspectratio,width=10cm]{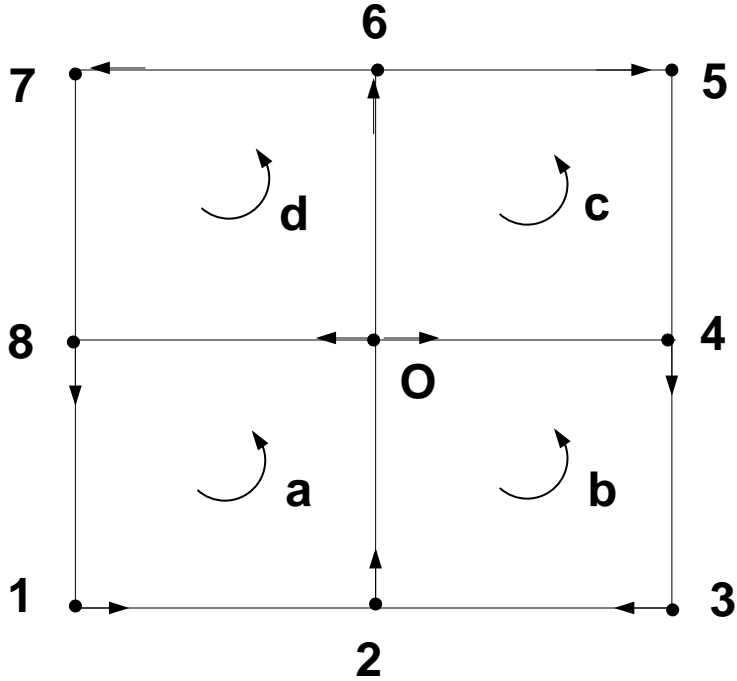}
\end{center}
\caption{2-vortex}
\end{figure}
 
  For a typical 2-vortex we refer to $Fig.2$ . We denote the subvortices by $\bf a$ , $\bf b$ , $\bf c$ and $\bf d$, each carrying topological charge $+1$ [ $Fig.2$ ] . However the spin at the central lattice point of the vortex ( the point O in $Fig.2$ ) becomes non-unique , as is clear from the construction of the subvortices . Thus the central point situated on a lattice site turns out to be a $\bf 'defect'$ or a singular point and cancellation of the contribution of ${{\mathcal S}_{WZ}^{stagg}}$ along the common bonds does not lead to a consistent spin field configuration . In other words the contributions of ${{\mathcal S}_{WZ}^{stagg}}$ on the boundary of the 2-vortex is not well defined  and the construction of 2-vortex in this scheme becomes problematic . The scenario persists in all the vortices (anti-vortices)  possessing $\bf even~~valued$ topological  charges , as can be read out from the spin field configuration in the case of 4-vortex $[Fig.4]$ (See Appendix ) . \\

{\bf 3.3 ~Analysis of 3-vortex }\\
\begin{figure}[!htbp]
\begin{center}
\includegraphics[keepaspectratio,width=10cm]{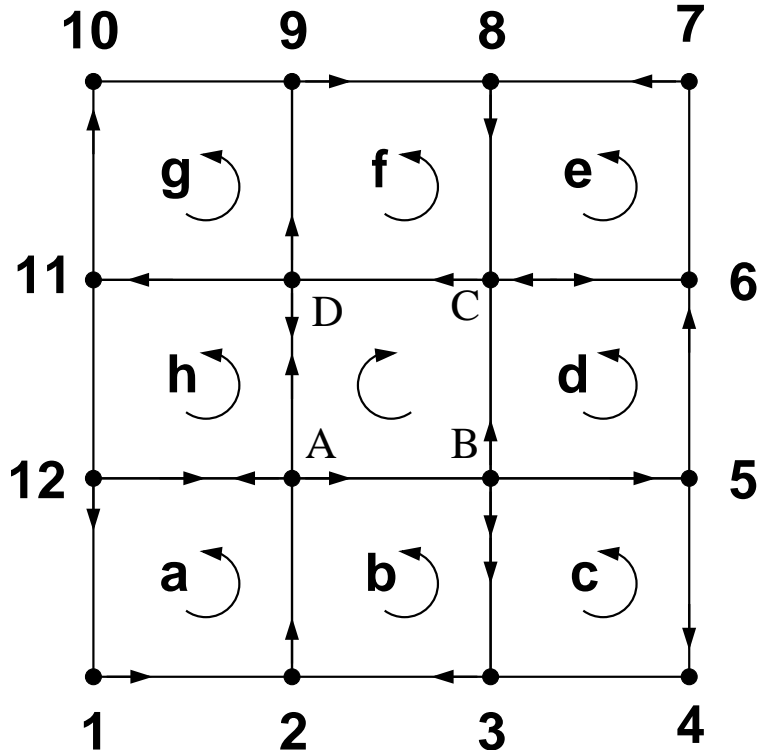}
\caption{3-vortex}
\end{center}
\end{figure}
 For 3-vortex  we have  a consistent spin field configuration [ $Fig.~3$] . We have now an elementary anti-vortex plaquette at the central region with well defined staggered spin field configurations   and  the contributions of ~${\mathcal S}_{WZ}^{stagg}$~ along the common bonds cancel each  other, giving rise to a  consistent spin field configurations .  It may be pointed out that in contrast to the case with even-valued charge  , we now have a subvortex with opposite charge occupying  the central region. \\

{\bf 3.4 ~Analysis of 4-vortex }\\
\begin{figure}[!htbp]
\begin{center}
\includegraphics[keepaspectratio,width=10cm]{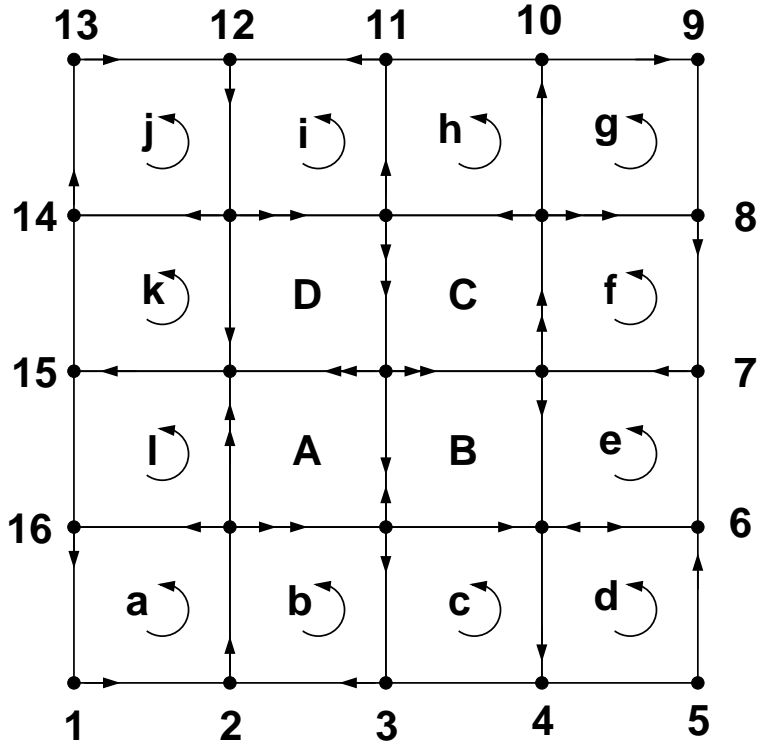}
\end{center}
\caption{4-vortex}
\end{figure}

 Similar to the situation corresponding to the 2-vortex , here again the spin at the centre becomes ambiguous, owing to the vanishing of the  effective horizontal component [ $Fig.~4$] . Thus the central point  turns out to  be a defect or a singular point . As a  result the contributions of  ~~${\mathcal S}_{WZ}^{stagg}$~~  on the common bonds do not cancel each other . The ambiguity of the suitable  spin field configuration  for 4-vortex is explained in the last paragraph of the appendix.\\

{\bf 4. ~Conclusion and Discussion}\\
i) Our calculations and analysis with $2D$ spin $\frac {1}{2}$ anti-ferromagnetic quantum $XY$ model on discrete lattice clearly brings out distingushing features between the even and odd charge sectors.  It may be recalled that the presence of a topological term in a field theoretic action indicates that the field configurations can be classified into different sectors with various topological charges . \\

ii) Our work has firmly established the role of WZ-like term , derived from a microscopic theory , as a topological charge indicator for the odd-charged excitations .  Our scheme is implemented in such a way that  only a net contribution along the outer boundary of a vortex plaquette remains nonvanishing . As a result ,  the WZ-like term  becomes an implicit function of the charge .\\

iii) Different  type of quantum mechanical operators were used to determine topological charges of excitations in spin models [14] . In our present work on QHAF we used a field theory based approach to identify such excitations.  \\

  Our future plan includes the generalization of our approach to the case of finite $\lambda_z$ to achieve a physical realization of meronic type of excitations in the spin models.  As  we  increase the magnitude of $\lambda_z$ from infinitesimal to a  finite value , the vortices/flattened merons would gradually go over to complete 3-dimensional merons . \\
Furthermore , we aim to evaluate the static and dynamic spin correlations for two dimensional spin $\frac {1}{2}$ anisotropic quantum Heisenberg anti-ferromagnet at any temperature  by taking into account the contributions from these topological excitations as well as spin waves.  These calculations will be very much useful to critically examine the novel concept of "quantum Kosterlitz - Thouless transition"  corresponding to quantum anti-ferromagnets [ 3, 15, 16, 17 ].  Above the Kosterlitz-Thouless transition temperature where the vortex-antivortex pairs unbind , the dynamical structure factor is expected to exhibit a central peak [6].\\

 Our picture has an interesting analogy with the work presented in [18].  However there are some finer differences. In our case we only have electrically neutral topological excitations which occur  at zero temperature as well as at finite temperature. On the other hand, in [18] there are both electrically charged and electrically neutral type of topological excitations. Whereas the charged ones appear at zero temperature, the neutral ones  appear only at finite temperature  . \\

 ~~~~To conclude, our study of topological spin excitations on the 2D-lattice will undoubtedly play an important role in the analysis of thermodynamics of low dimensional anti-ferromagnets.\\
{\bf Acknowledgement} : One of the authors (SKP) would like to thank M.G.Mustafa, Rajarshi Roy and Purnendu Chakraborti for their valuable help in the preparation of the figures.

\vspace{0.1cm}

{\bf Reference}\\

$1.$E. Fradkin and M. Stone , Phys. Rev. B {\bf 38}, 7215 (1988).\\

$2.$E. Fradkin, $Field$ $Theories$ $of$ $Condensed$ $Matter$ $Systems$ ( Addision-Wesley, CA, 1991).\\

$3.$Ranjan Chaudhury and Samir K. Paul , Phys. Rev. B {\bf 60} 6234 (1999).\\

$4.$E. Manousakis , Rev. Mod. Phys. {\bf 63} 1 (1991).\\

$5.$P. Weigmann , Phys. Rev. Lett. {\bf 60} 821 (1988).\\

$6.$Y. Endoh $et$ $al$., Phys. Rev. B $\bf 37$ 7443 (1988); K. Yamada $et$ $al.$, Phys. Rev. B $\bf 40$, 4557 (1989); R. Chaudhury , Indian J. Phys. A,$\bf 66$, 159 (1992).\\

$7.$F. G. Mertens, A. R. Bishop, G. M. Wysin and C. Kawabata  , Phys. Rev. Lett. {\bf 59} 117 (1987) ; D. L. Huber, Phys. Rev. {\bf B26} 3758 (1982) .\\ 

$8.$Ranjan Chaudhury and Samir K. Paul , Mod. Phys. Letts. B {\bf 16} 251 (2002).\\

$9.$Tai Kai Ng , Phys. Rev. Lett. {\bf 82} 3504 (1999) ; M. Greven $et$ $al.$  , Phys. Rev. Lett. {\bf 72} 1096 (1994) .\\

$10.$S. Chakraverty, B. F. Halperin and D. R. Nelson , Phys. Rev. Lett. {\bf 60} 1057 (1988) , S. Tyc, B.I. Halperin and S. Chakraverty, 62, 835 (1989)     \\

$11.$Mona Berciu and Sajeev John , Phys. Rev. B {\bf 61} 16454 (2000).\\

$12.$T. Morinari , J. Magn. Mag. Mat. {\bf 302} , 382 (2006).\\

$13.$ A. S. M\'ol , A. R. Pereira , H. Chamati and S. Romano , Eur. Phys. J. B {\bf 50} , 541 (2006).\\

$14.$E. Loh, Jr., D. J. Scalapino and P. M. Grant , Phys. Rev. B {\bf 31} 4712(1985) ;  R. H. Swendsen, Phys. Rev. Lett. {\bf 49}, 1302 (1982) ;  D. D. Betts, F. C. Salevsky and J. Rogiers, J. Phys. A {\bf 14} (1981); P. W. Anderson, Sajeev John, G. Baskaran, B. Doucot, S. D. Liang, Princeton University Preprint (unpublished , 1988).\\

$15.$A.Cuccoli , T. Valerio , V. Paola and V. Ruggero , Phys. Rev. B, {\bf 51} 12840 (1995).\\

$16.$J.M. Kosterlitz and D.J. Thouless, J. Phys. C, 6, 1181 (1973) ; Berezinskii V L , Sov. Phys. JETP ,{\bf 32} 493 (1970) ; ibid Sov. Phys. JETP ,{\bf 34} 610 (1972)  ; F. Fucito and S. Solomon, DOE Research and Development Report, CALT-68-1023 (1981) ; See also R. Chaudhury in [6] .\\

$17.$S. Komineas and N. Papanicolaou , Nonlinearity , {\bf 11} 265 (1998).\\

$18.$C. Timm and K. H. Benneman , Phy. Rev. Lett. {\bf 84} 4994 (2000). \\

{\bf {Appendix~~A}}

Applying variational  principle  to the action given by equation (10), we  obtain the equation of motion  for an XY-anisotropic Heisenberg antiferromagnet explicitly on the lattice , in the following form:
\begin{flushleft}
${{(-1)}^{i+j}}~{{\partial_t}{ n_1}}(ia,ja)  =  -  i~g~s~[{\lambda_z}{ n_2}{N_3} - { n_3}{N_2}](ia,ja)$
\end{flushleft}
\begin{flushleft}
${{(-1)}^{i+j}}~{{\partial_t}{ n_2}}(ia,ja)  =  -  i~g~s~[{ n_3}{N_1} - {\lambda_z}{ n_1}{N_3}](ia,ja)$
\end{flushleft}
\begin{flushleft}
${{(-1)}^{i+j}}~{{\partial_t}{n_3}}(ia,ja)  =  -  i~g~s~[{ n_1}{N_2} -{ n_2}{N_1}](ia,ja)$\hfill(A.1)\\
\end{flushleft}
In the above derivation we have used the following variation of the WZ-term given by [2]
\begin{flushleft}
$\delta {{\mathcal S}^{stagg}_{WZ}}[{\bf m}]  =  {\sum_{i,j}}~{{(-1)}^{i+j}}~\delta {{\mathcal S}_{WZ}}[{\bf m}(ia,ja)]$
\end{flushleft}
\begin{flushleft}
$=  {\sum_{i,j}}~{{(-1)}^{i+j}}~{\int_0^{\beta}}~dt~\delta {\bf n}(ia,ja)\cdot ({\bf n}\wedge {\partial_t}{\bf n})(ia,ja)$\hfill(A.2)\\
\end{flushleft}
In Eqn.(A.1) ${N_1}(ia,ja),{N_2}(ia,ja),{N_3}(ia,ja)$ are the components of the vector $\bf N$ at the lattice point $(ia,ja)$ . The vector $\bf N$ is given as:
\begin{flushleft}
${\bf N}(ia,ja) = {\bf n}(ia,(j-1)a) + {\bf n}((i-1)a,ja) + {\bf n}((i+1)a,ja) + {\bf n}(ia,(j+1)a)$\hfill(A.3)\\
\end{flushleft}
  From equations (11), (12) and (13)  we can write ${{\mathcal S}_{WZ}^{stagg}}$ in the following form :\\
\begin{flushleft}
$2{{\mathcal S}_{WZ}^{stagg}} =  2 {\sum_{i,j}}~{{(-1)}^{i+j}}~{{\mathcal S}_{WZ}}[{\bf n}(ia,ja)]$
\end{flushleft}
\begin{flushleft}
$ =  {\sum_{i,j}}~{{(-1)}^{i+j}}{\int_0^{\beta}}~dt~\{ [{\bf n}(i,j) - {\bf n}(i,j-1)]\cdot ({\bf n}\wedge {\partial_t}{\bf n})(i,j)$
\end{flushleft}
\begin{flushleft}
$ + [{\bf n}(i,j) - {\bf n}(i-1,j)]\cdot ({\bf n}\wedge {\partial_t}{\bf n})(i,j) - [{\bf n}(i,j-1) - {\bf n}(i-1,j-1)]\cdot ({\bf n}\wedge {\partial_t}{\bf n})(i,j-1) $
\end{flushleft}
\begin{flushleft}
$  - [{\bf n}(i-1,j) - {\bf n}(i-1,j-1)]\cdot ({\bf n}\wedge {\partial_t}{\bf n})(i-1,j) \}$\hfill(A.4)\\
\end{flushleft}

Thus for the 1-vortex ( anti-vortex ) plaquette having vertices 1 , 2 , 3 and 4 (see $Fig.1$ )   with coordinates $((i-1)a,(j-1)a)$, $(ia,(j-1)a)$, $(ia,ja)$ and $((i-1)a,ja)$ respectively  the contribution of  ${{\mathcal S}_{WZ}^{stagg}}$   to the 1-vortex    is  given by

\begin{flushleft}
$ {{[{{\mathcal S}_{WZ}^{stagg}}]}_{(1-vortex)}}  = -   {\int_0^{\beta}}dt [ {\bf n}((i-1)a,ja)\cdot ({\bf n}~\wedge  {{(-1)}^{i+j}}{\partial_t}{\bf n}) (ia,ja)$
\end{flushleft}
\begin{flushleft}
$ + {\bf n}(ia,(j-1)a)\cdot ({\bf n}\wedge {{(-1)}^{i+j}}{\partial_t}{\bf n}) (ia,ja) + {\bf n}((i-1)a,(j-1)a)\cdot ({\bf n}\wedge {{(-1)}^{i+j}}{\partial_t}{\bf n}) (ia,(j-1)a)$
\end{flushleft}
\begin{flushleft}
$+{\bf n}((i-1)a,(j-1)a)\cdot ({\bf n}\wedge  {{(-1)}^{i+j}}{\partial_t}{\bf n}) ((i-1)a,ja)] \}$\hfill(A.5)\\
\end{flushleft}
We evaluate the right hand side of $Eqn.(A.4)$ by substituting for  ${\partial_t}{\bf n}$ from  $Eqns.(A.1)~ and~ (A.3)$ . To keep the calculations  simple but consistent , we retain the intra-plaquette contributions by imposing a so called "local periodic boundary condition" (local PBC) as applied to the site closest to the vertices belonging to the plaquette under consideration .\\
For example, in the case of 1-vortex [see Fig.1a]  the local PBC implies:
\begin{flushleft}
${\bf n}(ia,ja) = {\bf n}((i-2)a,ja) = {\bf n}(ia,(j-2)a)$
\end{flushleft}
\begin{flushleft}
${\bf n}((i-1)a,ja) = {\bf n}((i-1)a,(j-2)a) = {\bf n}((i+1)a,ja)$
\end{flushleft}
\begin{flushleft}
${\bf n}(ia,(j-1)a) = {\bf n}(ia,(j+1)a) = {\bf n}((i-2)a,(j-1)a)$
\end{flushleft}
\begin{flushleft}
${\bf n}((i-1)a,(j-1)a) = {\bf n}((i+1)a,(j-1)a) = {\bf n}((i-1)a,(j+1)a)$\hfill(A.6)\\
\end{flushleft}
 Besides we make use of  the following conditions satisfied  at different vertices of the plaquette for all time, as explained in the section 3.1
 [see  Fig.1a] :
\begin{flushleft}
${ n_1}((i-1)a,(j-1)a) = - { n_1}(ia,ja) = 1 - \delta $
\end{flushleft}
and
\begin{flushleft}
${ n_2}(ia,(j-1)a) = - { n_2}((i-1)a,ja) = 1 - \delta$\hfill(A.7)\\
\end{flushleft}
This means that the following equations must hold :
\begin{flushleft}
${\partial_t}{ n_1}((i-1)a,(j-1)a)  =  - {\partial_t}{ n_1}(ia,ja)  =  {\partial_t}(1 - \delta )$
\end{flushleft}
\begin{flushleft}
${\partial_t}{ n_2}(ia,(j-1)a)  =  - {\partial_t}{ n_2}((i-1)a,ja)  =  {\partial_t}(1 - \delta )$\hfill(A.8)\\
\end{flushleft}
 Making use of Eqns. (A.3) and  (A.6) in (A.1) we have
\begin{flushleft}
$ {{(-1)}^{i-1+j-1}} {{\partial_t}{ n_1}}((i-1)a,(j-1)a)  =  -i~g~s~{\lambda_z}{ n_2}{N_3}((i-1)a,(j-1)a) $
\end{flushleft}
\begin{flushleft}
${{(-1)}^{i+j}}  {{\partial_t}{ n_1}}(ia,ja)  =  -  i~g~s~{\lambda_z}{ n_2}{N_3}(ia,ja)$
\end{flushleft}
\begin{flushleft}
${{(-1)}^{i+j-1}}   {{\partial_t}{ n_2}}(ia,(j-1)a)  =  i~g~s~{\lambda_z}{ n_1}{N_3}(ia,(j-1)a)$
\end{flushleft}
\begin{flushleft}
${{(-1)}^{i-1+j}}   {{\partial_t}{ n_2}}((i-1)a,ja)  =  i~g~s~{\lambda_z}{ n_1}{N_3}((i-1)a,ja)$ \hfill(A.9)\\
\end{flushleft}
 The right hand sides of Eqs. (A.9) vanish in the flattened meron configuration limit where ${\lambda_z}\longrightarrow 0$  and   ${N_3}$  acquires very small value , since ${\epsilon}(ia,ja)$ is very small in this configuration. Consequently  from Eqns.(A.8) it follows that the quantity $ \delta $  does not vary with time . Note that  $\delta $  is independent of the position of the vertex points of the plaquette [See Fig.1.a] , as explained in section 3.1 \\
   It is important and interesting to point out that had we not used the local PBC (A.6), we would not have had "time independent $\delta$" . In other words, the spin field configurations forming the vortices become non-stationary in absence of the local PBC .  For example using  the 1st equation under (A.1) and the equation (A.3) we have :\\ 
$ {{(-1)}^{i-1+j-1}} {{\partial_t}{ n_1}}((i-1)a,(j-1)a)  =  -i~g~s~[{\lambda_z}{ n_2}{N_3}((i-1)a,(j-1)a)-{ n_2}((i-1)a,(j-2)a)-{ n_2}((i-2)a,(j-1)a)]$ \\
 The right hand side ( rhs ) of the above equation does not vanish even in the flattened meron configuration limit. This is unlike in the case of the 1st equation under (A.9) where  the parameter  $\delta $ came out to be time independent. However application  of the local PBC (A.6) restores the rhs (as referred to earlier) to that of the 1st equation under (A.9).\\       
  
   Now using the Eqns. (A.1), (A.3),  (A.5), (A.6) and (A.7)  we obtain ~${\mathcal S}_{WZ}$~   as  follows :
\begin{flushleft}
$ {{[{{\mathcal S}_{WZ}^{stagg}}]}_{(1-vortex)}}   =  [{n_1}(2)+{n_1}(4)][({\bf n}\cdot {\bf N})(3) {n_1}(3) - {N_1}(3)]$
\end{flushleft}
\begin{flushleft}
$+[{n_3}(2)+{n_3}(4)][({\bf n}\cdot {\bf N})(3) {n_3}(3) - {N_3}(3)] + {n_1}(1)[({\bf n}\cdot {\bf N})(4){n_1}(4) + ({\bf n}\cdot {\bf N})(2){n_1}(2)]$
\end{flushleft}
\begin{flushleft}
$ + {n_2}(1)[({\bf n}\cdot {\bf N})(4){n_2}(4) + ({\bf n}\cdot {\bf N})(2){n_2}(2) - 2{ N_2}(2)] + {n_3}(1)[({\bf n}\cdot {\bf N})(4){n_3}(4) $
\end{flushleft}
\begin{flushleft}
$+ ({\bf n}\cdot {\bf N})(2){n_3}(2) - 2{ N_3}(2)]$
\end{flushleft}
\begin{flushleft}
$\equiv  A + B$\hfill(A.10)\\
\end{flushleft}

where in the flattened meron configuration limit $A$  and  $B$  are  given  by:
\begin{flushleft}
$A  = {\int_0^{\beta}}dt~~ \{  2{{({n_1}(2)+{n_1}(4))}^2}[{{n_1}^2}(3) - 1] + 2~{n_1}(3)~{n_3}(3)~[{n_1}(2)+{n_1}(4)][{n_3}(2)+{n_3}(4)]$
\end{flushleft}
\begin{flushleft}
$ + 4~{n_2}(1)~[{{n_2}^2}(2) - 1][{n_2}(3)+{n_2}(1)]   \} $
\end{flushleft}
\begin{flushleft}
$B  = {\int_0^{\beta}}dt~~ \{   + 2~{n_2}(2)~{n_1}(1)~[{n_1}(2)-{n_1}(4)]~[{n_2}(3)+{n_2}(1)]$
\end{flushleft}
\begin{flushleft}
$ + 2~{n_2}(2)~{n_3}(1)~[{n_3}(2)-{n_3}(4)]~[{n_2}(3)+{n_2}(1)] \}$\hfill(A.11)\\\end{flushleft}

 In the equations (A.10) and (A.11) above we have labelled the vertices of the 1-vortex [Fig.1a] by 1, 2, 3 and 4 in the anti-clockwise sense. The quantity ${n_i}$(1) denotes the i-th component (i = 1, 2, 3) of the coherent spin vector $\bf n$ at the vertex 1 and similarly  ${n_i}$(2), ${n_i}$(3) and  ${n_i}$(4) stand for the i-th components at the vertices 2, 3 and 4 respectively .\\ 
It is interesting to note that $A$ remains invariant if we go from vortex to antivortex by changing  ${n_2}(2)$ and ${n_2}(4)$ in $Fig.1$ to $-{n_2}(2)$ and $-{n_2}(4)$ respectively whereas $B$ goes over to $-B$ . Thus  ~${[{{\mathcal S}_{WZ}^{stagg}}]}_{(1-vortex)}$~  takes the form $A-B$ for antivortex $[Fig1.b]$.\\

To analyse the case of 2- vortex [Fig.2] we assign the co-ordinates to the vertices 1, 2, 3, 4, 5, 6, 7, 8 as $((i-1)a,(j-1)a)$ , $(ia,(j-1)a)$, $((i+1)a,(j-1)a)$, $((i+1)a,ja)$, $((i+1)a,(j+1)a)$, $(ia,(j+1)a)$, $((i-1)a,(j+1)a)$, $((i-1)a,ja)$ respectively .  The centre O  has co-ordinates  $(ia,ja)$. Using Eqns. (A.1) , (A.3) as well as the local PBC , as appropriate to the case of 2-vortex , the equations of motion  of the spins ${n_2}$ at the vertices  2, 4, 6, 8 reduce to the following , in the flattened  meron  configuration limit ( ${{\lambda}_z}\longrightarrow 0$ ):
\begin{flushleft}
${{(1-\delta )}^2} = {{cos}^2}{\epsilon (2)} = {{cos}^2}{\epsilon (4)} = {{cos}^2}{\epsilon (6)} = {{cos}^2}{\epsilon (8)}$\hfill(A.12)\\
\end{flushleft}
It is therefore obvious from the above equations that the z-component of the spin $\bf n$ viz., $n_3$ becomes position independent.  \\
 In Fig.3 we have denoted the elementary vortex plaquettes by $\bf a$, $\bf b$, $\bf c$, $\bf d$, $\bf e$, $\bf f$, $\bf g$, $\bf h$. Note that there is an elementary anti-vortex plaquette in the central region with vertices $\bf A$, $\bf B$, $\bf C$ and $\bf D$ .  We write formally the expression for ~${{\mathcal S}_{WZ}^{stagg}}_{(3-vortex)}$~ :
\begin{flushleft}
~${{[{{\mathcal S}_{WZ}^{stagg}}]}_{(3-vortex)}}  =   {\int_0^{\beta}}dt   \{  {\bf n}(1a)\cdot ({\bf n}\wedge {\partial_t}{\bf n})(2a) +{\frac{1}{2}}{\bf n}(2a)\cdot ({\bf n}\wedge {\partial_t}{\bf n})(3a) +{\frac{1}{2}}{\bf n}(4a)\cdot ({\bf n}\wedge {\partial_t}{\bf n})(3a) + {\bf n}(1a)\cdot ({\bf n}\wedge {\partial_t}{\bf n})(4a)$
\end{flushleft}
\begin{flushleft}
$ + {\bf n}(1b)\cdot ({\bf n}\wedge {\partial_t}{\bf n})(2b) + {\bf n}(2b)\cdot ({\bf n}\wedge {\partial_t}{\bf n})(3b)+{\frac{1}{2}}{\bf n}(4b)\cdot ({\bf n}\wedge {\partial_t}{\bf n})(3b) +{\frac{1}{2}}{\bf n}(1b)\cdot ({\bf n}\wedge {\partial_t}{\bf n})(4b)$
\end{flushleft}
\begin{flushleft} 
$ +{\frac{1}{2}}{\bf n}(1c)\cdot ({\bf n}\wedge {\partial_t}{\bf n})(2c) + {\bf n}(2c)\cdot ({\bf n}\wedge {\partial_t}{\bf n})(3c)+ {\bf n}(4c)\cdot ({\bf n}\wedge {\partial_t}{\bf n})(3c) +{\frac{1}{2}}{\bf n}(1c)\cdot ({\bf n}\wedge {\partial_t}{\bf n})(4c)$
\end{flushleft}
\begin{flushleft}
$+ ..............................$
\end{flushleft}
\begin{flushleft}
$+{\frac{1}{2}}{\bf n}(1h)\cdot ({\bf n}\wedge {\partial_t}{\bf n})(2h) +{\frac{1}{2}}{\bf n}(2h)\cdot ({\bf n}\wedge {\partial_t}{\bf n})(3h)+{\frac{1}{2}} {\bf n}(4h)\cdot ({\bf n}\wedge {\partial_t}{\bf n})(3h) + {\bf n}(1h)\cdot ({\bf n}\wedge {\partial_t}{\bf n})(4h)  \}$\hfill(A.13)\\
\end{flushleft}
In the above equation 1a, 2a, 3a and 4a denote the vertices for the subvortex $\bf a$ in the anti-clockwise sense , as we already have  in the case of 1-vortex. Similarly for the other subvortices. It can be shown after  a long calculation  that Eqn. (A.13)  can be written in the form  $A + B$ , as we had in the case of 1-vortex , where $B$  consists of terms  linear  in ${ n_2}(2)$ or containing odd powers of  ${ n_2}(2)$. Thus $A + B$  goes over to $A - B$ as we  go from vortex to anti-vortex.\\
In the case of a 4-votex [Fig.4] we have a again a single lattice point as in the case of the 2-vortex. By similar reasons the 4-vortex construction in terms of elementary plaquettes, breaks down in the flattened meron configuration limit.

\end{document}